\documentclass[showpacs,amsmath,amssymb,aps,showkeys,floatfix,prd,a4paper]{revtex4}
\usepackage[dvips]{graphicx}
\usepackage{dcolumn}
\usepackage{bm} 
\usepackage{epsfig}
\usepackage{amsfonts}
\usepackage{amssymb,amscd}
\usepackage{subfigure}
\usepackage{xcolor}
\usepackage{verbatim}

\newcommand{\ben}{\begin{eqnarray}}
\newcommand{\een}{\end{eqnarray}}

\newcommand{\IFUSP}{Instituto de F\'{\i}sica, Universidade de S\~{a}o Paulo,
Rua do Mat\~ao, 1371,  05508-090, \\   Cidade Universit\'aria, S\~{a}o Paulo, SP, Brazil.}

\newcommand{\salvador}{ 
Instituto de F\'{\i}sica, Universidade Federal da Bahia,
Campus Universit\'ario de Ondina, 40170-115, Bahia, Brazil.}

\newcommand{\unifesp}{
Universidade Federal de S\~ao Paulo, C.P. 01302-907, S\~ao Paulo, Brazil.}

\begin{document}

\title{New charm resonances and $J/\psi$ interactions in a hadron gas}

\author{L.~M. Abreu$^{1}$}
\author{K.~P.~Khemchandani$^{2}$}
\author{A.~Mart\'{\i}nez~Torres$^{3}$}
\author{F.~S. Navarra$^{3}$}
\author{M.~Nielsen$^{3}$}
\affiliation{$^{1}$ \salvador}
\affiliation{$^{2}$ \unifesp}
\affiliation{$^{3}$ \IFUSP}

\begin{abstract}

In relativistic heavy ion collisions, after the quark gluon plasma (QGP) phase there is a 
hadron gas (HG) phase. In both phases $J/\psi$ may be formed and destroyed.        
In this note we study the $J/\psi$ interactions with other mesons in 
the hadron gas 
phase. Making use of effective field Lagrangians we obtain the cross sections 
for the production and absorption processes.  
With respect to the existing calculations, the improvements introduced here
are the inclusion of $K$ and $K^*$'s in the effective Lagrangian approach
(and the computation of the corresponding cross sections)
and the inclusion of processes involving the new exotic charmonium states
$Z_c(3900)$ and $Z_c(4025)$. 

We conclude that the interactions between $J/\psi$ and  all the
considered  mesons reduce the original $J/\psi$ abundance (
determined at the end of the quark gluon plasma phase) by  20 \% and 
24 \% in RHIC and LHC  collisions respectively.  Consequently, any really
significant change in the $J/\psi$  abundance comes from  dissociation and
regeneration processes in the QGP phase.

\end{abstract}


\maketitle

\section{Introduction}

Quark gluon plasma (QGP) is formed in relativistic heavy ion collisons. 
In the QGP phase $J/\psi$'s are destroyed
and created in a complex and rich dynamical process, which involves many
properties of the QGP which we wish to know better \cite{qgp}. After cooling 
and hadronization there is a hadron gas (HG) phase, in which the $J/\psi$'s
can be destroyed and created in interactions with other mesons.   
New charmonium states, the so-called X,Y and Z  states \cite{review}, open  
new  channels for the $J/\psi$-light meson reactions, as first noticed in Ref. 
\cite{brazzi}.
In \cite{nos}  the existing calculations of the  $J/\psi$ - strange
meson dissociation cross sections were improved and also the 
cross sections of processes involving $Z_c(3900)$ 
and $Z_c(4025)$ intermediate states were calculated for the first time. 
Here we present a summary of \cite{nos} and briefly discuss some questions raised 
after its appearance. 

\section{$J/\psi$ - meson interactions}

The $J/\psi$ production processes 
$D^{(\ast)} + \bar{D}^{(\ast)} \to J/\psi + \pi $, 
$D^{(\ast)} + \bar{D}^{(\ast)} \to J/\psi + \rho $,
$D_s ^{(\ast)} + \bar{D} ^{(\ast)} \to J/\psi + K $,
$D_s ^{(\ast)} + \bar{D} ^{(\ast)} \to J/\psi + K^* $ 
can be studied with mesonic effective Lagrangians \cite{lee}.  
In \cite{nos} these
interactions were treated within the framework of an $SU(4)$ effective 
theory. The relevant Lagrangians are given by \cite{lee,nos}
\begin{eqnarray}
\mathcal{L}_{PPV} & = & -ig_{PPV}\langle V^\mu[P,\partial_\mu P]\rangle ,  
\nonumber \\
\mathcal{L}_{VVV} & = & i g_{VVV} \langle \partial_\mu V_\nu \left[ V^{\mu}, V^{\nu} 
\right] \rangle , 
\nonumber \\
\mathcal{L}_{PPVV} & = & g_{PPVV}\langle P V^\mu[V_\mu , P]\rangle ,  
\nonumber \\
\mathcal{L}_{VVVV} & = & g_{VVVV}\langle V^\mu V^\nu [V_\mu , V_\nu]\rangle , 
\label{Lagr1}
\end{eqnarray}
where the indices $PPV$ and $VVV$, $PPVV$ and $VVVV$ denote the type of vertex   
incorporating pseudoscalar and vector meson fields in the couplings  and 
$g_{PPV}$, $g_{VVV}$, $g_{PPVV}$ and $g_{VVVV}$ are the respective 
coupling constants. The symbol $\langle \ldots \rangle$ stands for the 
trace over $SU(4)$-matrices. The symbols $V_\mu$ and $P$ represent the 
corresponding  $SU(4)$ 15-plets of vector  and pseudoscalar fields respectively
(see \cite{nos} and references therein for details).
It is important to include the  anomalous parity terms:
\begin{eqnarray} 
\mathcal{L}_{PVV} & = & - g_{PVV} \varepsilon^{\mu\nu\alpha\beta} \langle 
\partial_\mu V_\nu \partial_\alpha V_\beta P \rangle ,  \nonumber \\
\mathcal{L}_{PPPV} & = & - i g_{PPPV} \varepsilon^{\mu\nu\alpha\beta} 
\langle V_\mu (\partial_{\nu} P) (\partial_{\alpha} P) (\partial_{\beta} P) 
\rangle , 
\nonumber \\
\mathcal{L}_{PVVV} & = & i g_{PVVV} \varepsilon^{\mu\nu\alpha\beta} \left[  
\langle V_\mu  V_\nu  V_\alpha \partial_{\beta} P \rangle   
+  \frac{1}{3} \langle V_\mu (\partial_\nu V_\alpha)  V_\beta P 
\rangle 
\right].
\label{Lagr2}
\end{eqnarray}
The $g_{PVV}$, $g_{PPPV}$, $g_{PVVV}$ are the coupling constants of the 
$PVV$, $PPPV$ and $PVVV$ vertices. As usual,  form factors are included in  
the vertices to account for higher order corrections and finite sizes effects.  
These form factors  can be calculated in different approaches (see, for example,  
\cite{ff}) but very often one uses  simple  parametrizations, which contain a 
cut-off parameter. In \cite{nos} we have used expressions of the type
$ F = (\Lambda^2) /  (\Lambda ^2 + \mathbf{q} ^2)$, where $\mathbf{q}$ is 
the three-momentum flowing in the vertex and $\Lambda$ is a cut-off mass of the
order of $2$ GeV.

\section{The effect of the new charmonium states}

The  state $Z_c(3900)$ opens a  new s-channel for $J/\psi$ interactions: 
$J/\psi +  \pi \to  Z_c \to D + \bar{D}^* $. 
The amplitude for this process can be written as
\begin{equation}
\mathcal{M}_{Z} =
\frac{ \alpha_{J/\psi \pi} \, \alpha_{D\bar D^*}}{s-M^2_Z+i M_Z\Gamma_Z} 
\times \, \left(-g^{\mu\nu}+\frac{p^\mu k^{\,\prime\nu}}{M^2_Z}\right)
\epsilon^\mu(k) \epsilon^{\nu\,*}(p^\prime) \, , 
\label{MZ}
\end{equation}
where $M_Z $  and $\Gamma_Z $  represent the mass and width  
of the $Z_c(3900)$ respectively. Also,   $\alpha_{J/\psi\pi}$ and  
$\alpha_{D\bar D^*}$ are the couplings of the $Z_c$ to the $J/\psi \pi$ and to    
the $D\bar D^{*}$ states, respectively.

The  state $Z_c(4025)$, which is assumed to be a $I^G(J^{PC})=1^+(2^{++})$ 
resonance, opens another  s-channel for $J/\psi$ interactions:
$ J/\psi + \rho \to Z_c(4025) \to D^* + \bar{D}^*$. 
The amplitude of this  process is given by:
\begin{equation}
\mathcal{M}_{Z^\prime} = 
\frac{ \eta_{J/\psi\rho} \, \eta_{D^*\bar D^*}}{s-M^2_{Z^\prime}+
i M_{Z^\prime}\Gamma_{Z^\prime}}  
\times \, P^{\mu\nu\alpha\beta}(q)
\epsilon_\mu(k)\epsilon_\nu(p)\epsilon^*_\alpha(k^\prime)
\epsilon^*_\beta (p^\prime),
\end{equation}
where $M_{Z^\prime}$ and $\Gamma_{Z^\prime}$  are  the mass and width 
of the $Z_c(4025)$ respectively. $\eta_{J/\psi\rho}$ and $\eta_{D^*\bar D^*}$ 
are the couplings of $Z_c(4025)$ to the channels $J/\psi \rho$ and $D^*\bar D^*$
and $P^{\mu\nu\alpha\beta}(q)$ is the spin 2 projector.  
The coupling constants $\eta_{J/\psi\rho}$ and $\eta_{D^*\bar D^*}$
were estimated in previous works. 

\begin{figure}[t]
  \begin{tabular}{cc}
   {\psfig{figure=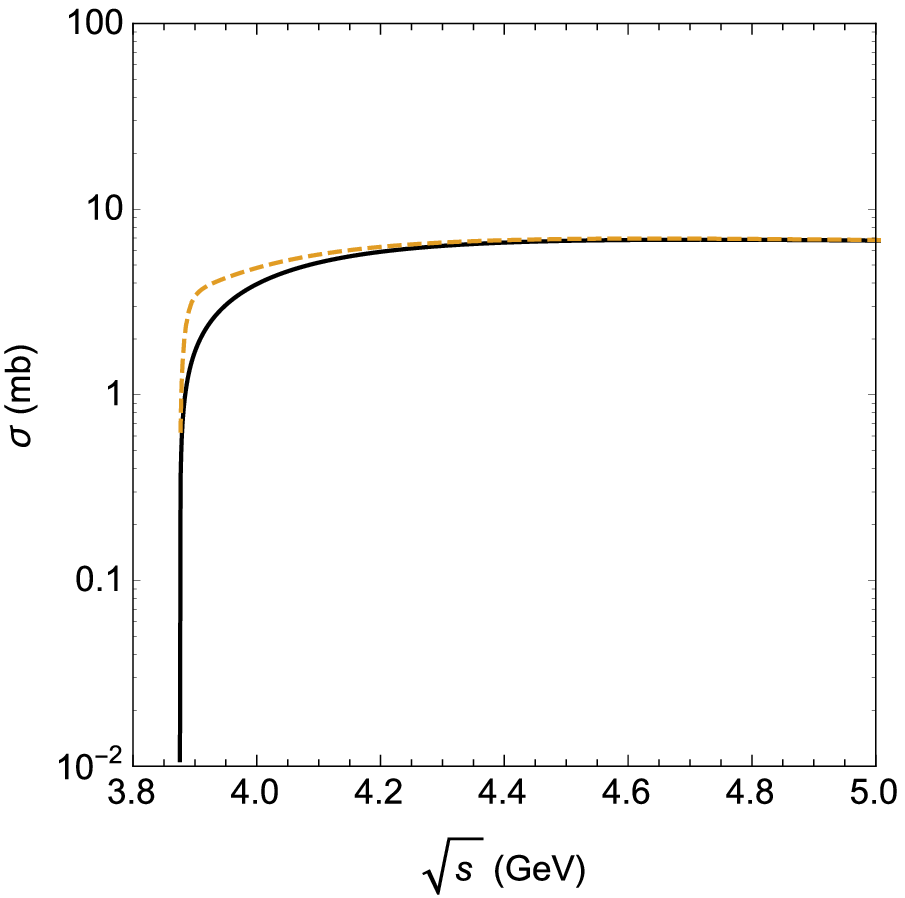,scale=0.75}} &
   {\psfig{figure=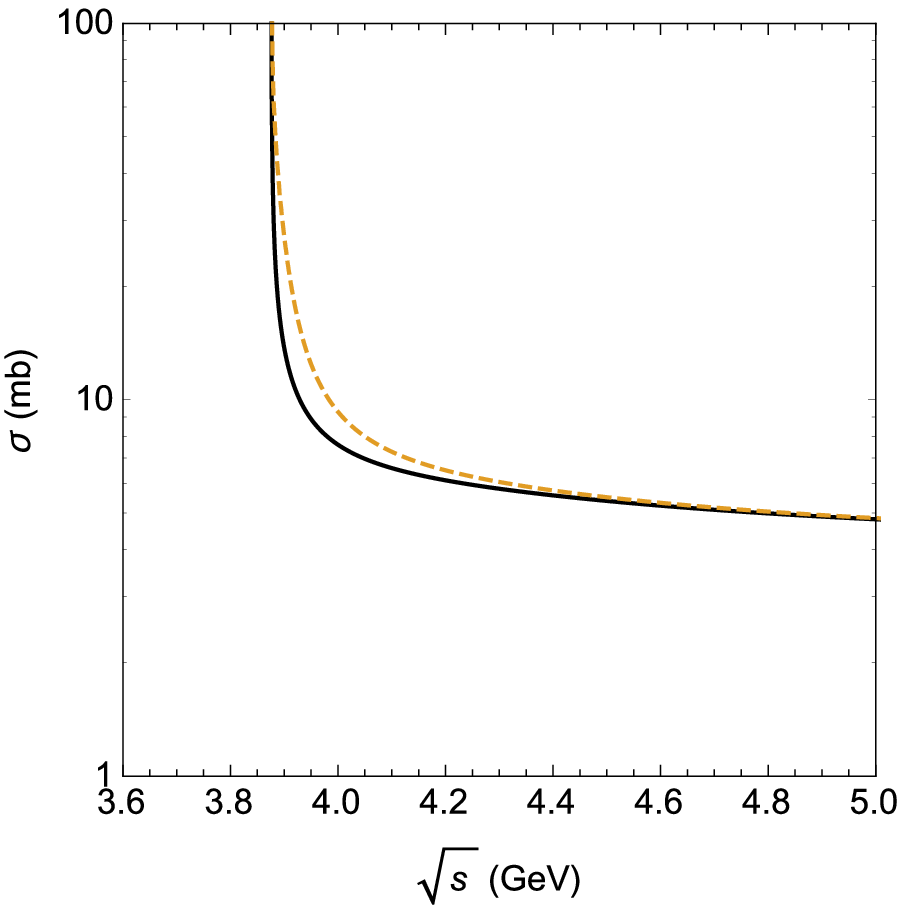,scale=0.75}}  \\
   {\psfig{figure=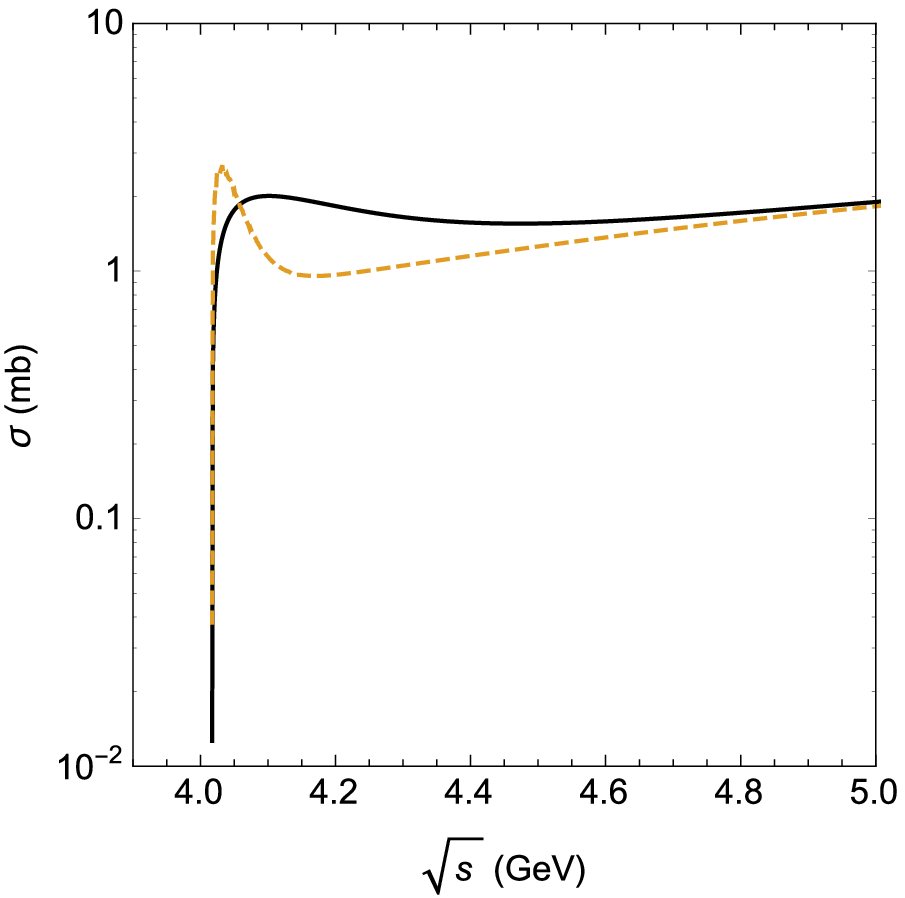,scale=0.75}} &
   {\psfig{figure=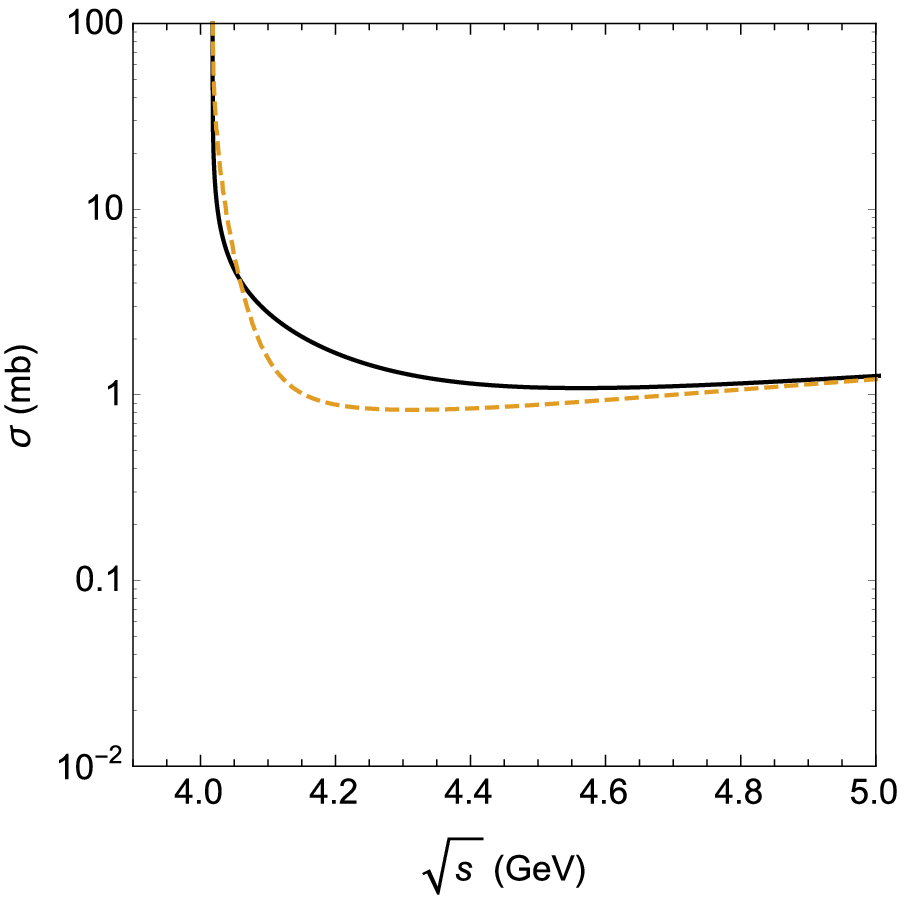,scale=0.75}}
  \end{tabular}
\caption{$J/\Psi$ absorption (top left) and production (top right ) cross sections
by $\pi$'s. The solid lines represent the cross sections obtained without including
the $Z_c$ (3900) exchange in the s-channel. The dashed lines show the results with
the exchange of $Z_c$ (3900) in the s-channel included. Bottom panels show the
$J/\Psi$ absorption (bottom left ) and production (bottom right ) cross sections
by $\rho$ ’s. The solid lines in these panels show the cross sections obtained
without  including the $Z_c$ (4025) exchange in the s-channel. The dashed lines
show the results obtained by including the $Z_c$(4025) exchange in the s-channel.
}
\label{fig1}
\end{figure}

In Fig.~\ref{fig1}  we show the cross sections of the  processes  
$J/\psi + \pi \,  \to  (Z_c(3900))  \to D + \bar{D}^*$  (top left),   
$D + \bar{D}^*  \, \to (Z_c(3900))  \to J/\psi + \pi$  (top right), 
$J/\psi + \rho \,  \to ( Z_c(4025))  \to  D^* + \bar{D}^*$ (bottom left) and 
$D^* + \bar{D}^* \, \to (Z_c(4025))  \to J/\psi + \rho$ (bottom right). 
The solid lines show the 
results obtained without the new resonances and the dashed lines show the 
effect of including the $Z_c(3900)$ and $Z_c(4025)$. As can be seen the 
effect of the new resonances is  small. 

\section{Time evolution of the $J/\psi$ abundance}

For each cross section mentioned above we can compute the 
thermally averaged cross section, which for  a given process $a b \rightarrow c d$ is 
given by 
\begin{equation}
\Sigma_{ab \to cd} =  
\frac{ 
\int \, d^{3} \mathbf{p}_a \, d^{3} \mathbf{p}_b \,  f_a(\mathbf{p}_a) \,  f_b(\mathbf{p}_b) \,  
\sigma_{a b \rightarrow c d } \, \,v_{a b} }
{\int \, d^{3} \mathbf{p}_a \, d^{3} \mathbf{p}_b \, f_a(\mathbf{p}_a) \,  f_b(\mathbf{p}_b) } 
\label{thermavcs}
\end{equation}
where $v_{ab}$ represents the relative velocity of initial two interacting 
particles $a$ and $b$ and the function $f_i(\mathbf{p}_i)$ is the 
Bose-Einstein distribution (of particles of species i), which depends on 
the temperature $T$.  With the help of (\ref{thermavcs}) we can study the time evolution of the $J / \psi$ abundance in 
a hot hadronic medium. The momentum-integrated  evolution equation has the form: 
\ben
\frac{d N_{J/\psi} }{d \tau} & = &  
\sum_{a, b} \sum_{\varphi } 
\Sigma_{a \, b \to \varphi \, J/\psi} \, n_{a} \,  N_{b}  \,  - 
\Sigma_{\varphi \, J/\psi \to a \, b} \, n_{\varphi} \,  N_{J/\psi}  
\label{rateeq}
\een
where $n_{\varphi} (\tau)$ are $N_{\varphi}(\tau)$ denote  the density and 
the abundances of  $\pi,\rho, K, K^{\ast}$, charmed mesons and their 
antiparticles in hadronic matter at  proper time $\tau$. $a$ and $b$ are $D$,
$D^*$, $D_s$ and $D_s^*$ mesons and their antiparticles.

We  assume that $\pi, \rho , K, K^{\ast}, D$ and  
$D^{\ast}$ are in equilibrium. Therefore the density $n_{i} (\tau)$ can be 
written as
\ben n_{i} (\tau) &  \approx & \frac{1}{2 \pi^2}\gamma_{i} g_{i} m_{i}^2 
T(\tau)K_{2}\left(\frac{m_{i} }{T(\tau)}\right), 
\label{densities}
\een
where $\gamma _i$ and $g_i$ are respectively the fugacity factor and the  
degeneracy factor of the relevant particle. The abundance $N_i (\tau)$ is 
obtained by multiplying  the density $n_i(\tau)$ by the volume $V(\tau)$. 
The time dependence is introduced through the temperature $T(\tau)$ and 
volume $V(\tau)$ profiles appropriate to model the dynamics of relativistic 
heavy ion collisions after the end of the quark-gluon plasma phase. The 
hydrodynamical expansion and cooling of the hadron gas is modeled by 
the boost invariant Bjorken flow  with an accelerated transverse expansion:
$$
T(\tau)  =  T_C - \left( T_H - T_F \right) \left( \frac{\tau - \tau _H }
{\tau _F -  \tau _H}\right)^{\frac{4}{5}} \hspace{0.5cm}
V(\tau)  =  \pi \left[ R_C + v_C 
\left(\tau - \tau_C\right) + \frac{a_C}{2} \left(\tau - \tau_C\right)^2 
\right]^2 \tau_C.
$$
In the equation above, $R_C $ and $\tau_C$  denote the final 
transverse  and longitudinal sizes of the quark-gluon plasma, while $v_C $ 
and  $a_C $ are its transverse flow velocity and transverse            
acceleration at this time. $T_C = 175$ MeV is the critical temperature     
for the quark-gluon plasma to hadronic matter transition; 
$T_H = T_C = 175$ MeV  is 
the temperature of the hadronic matter at the end of the mixed phase,    
occurring at the time $\tau_H $. The freeze-out temperature  $T_F = 125$ 
MeV  then  leads to a freeze-out time $\tau _F $. 
We show results for the $J/\psi$ evolution in the hadron gas formed in two types  
of collisions: central $Au - Au$ collisions at $\sqrt{s_{NN}}= 200$ GeV 
at RHIC and central $Pb-Pb$ collisions at $\sqrt{s_{NN}} = 5$ TeV at the LHC. 
The parameters which we need as input are taken from the phenomenological 
studies of  the EXHIC Collaboration \cite{exhic}.

\begin{widetext}
\begin{figure}[th]
\centering
\includegraphics[width=6.5cm]{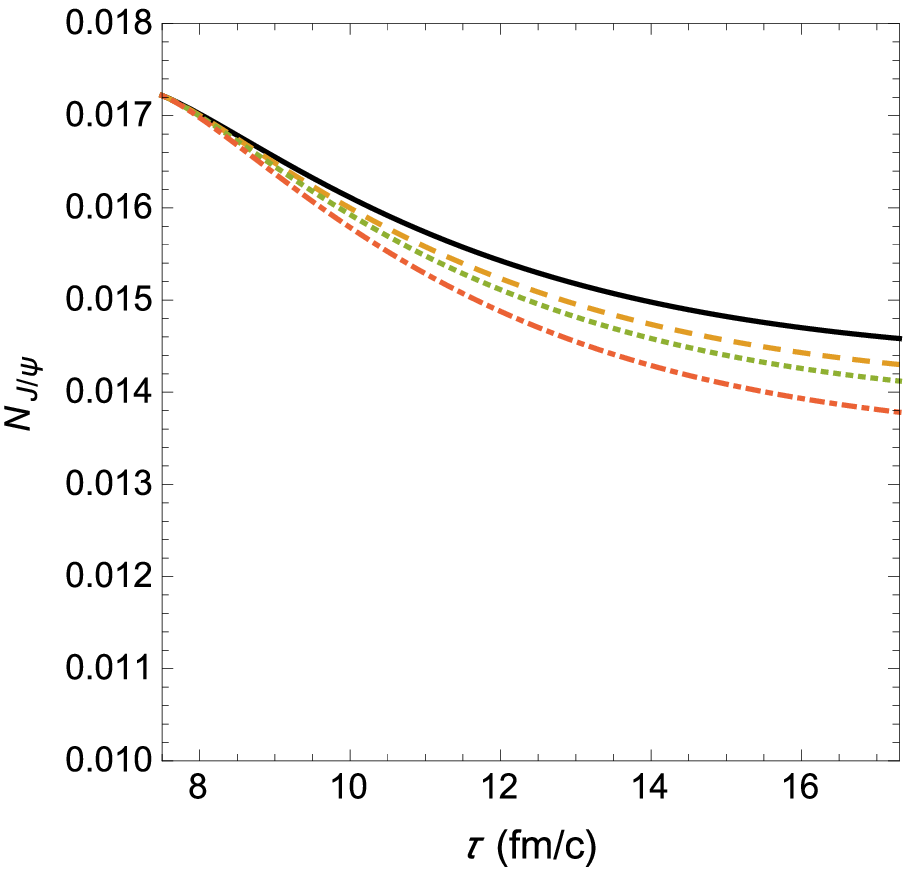}
\hspace{0.6cm}
\includegraphics[width=6.5cm]{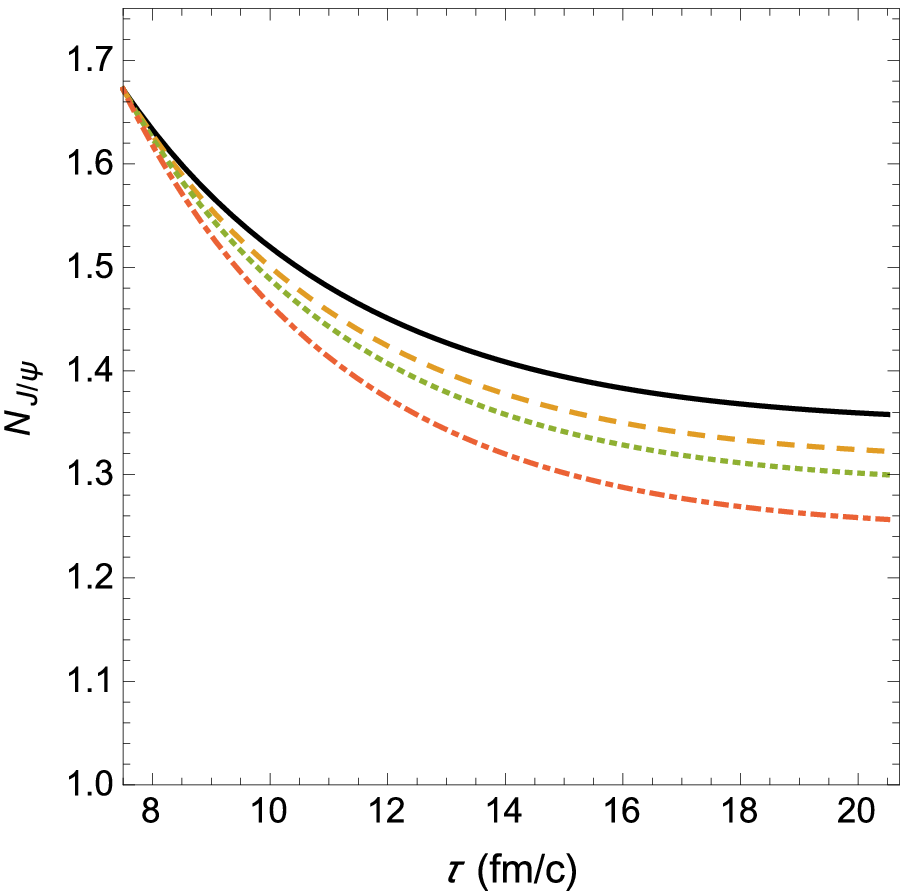}
\caption{ Left: Time evolution of $J/\psi$ abundance as a function of the proper
time in  central Au-Au collisions at $\sqrt{s_{NN}} = 200$ GeV. Solid, dashed,
dotted, dot-dashed  lines represent the situations with only $ \pi - J/\psi$
interactions and also adding the  $\rho - J/\psi$, $ K - J/\psi$ and
$ K^{\ast} - J/\psi$ contributions, respectively.
Right: the same as on the left for LHC conditions.
}
\label{fig2}
\end{figure}
 \end{widetext}

In Fig. \ref{fig2} we present the time evolution of the $J/\psi$ 
abundance as a function of the proper time for the two types of collisions 
discussed above: at RHIC (on the left panel) and at the LHC (on the right 
panel).  The different curves represent the interactions: 
only $J/\psi \, \pi$ (solid lines); adding $J/\psi \, \rho$ (dashed lines); 
adding  $J/\psi \, K$ (dotted lines) and adding also  $J/\psi \, K^*$  
(dash-dotted lines). 

While the cross sections alone would lead to an enhancement of
the $J/\psi$ yield, the relative multiplicities favor its
reduction, since in the hadron gas there are much more pions and kaons
(which hit and destroy the charmonium states) than $D$'s, $\bar{D}$'s,
$D_s$'s and $\bar{D_s}$'s (which can collide and create them). The result
of this  competition is a  decrease of the $J/\psi$ yield of approximately
20 \% at RHIC and 24 \% at the LHC.

\section{Summary}

In this note we have described how to improve the study of  
$J/\psi$ interactions in a hadron gas 
(in the effective Lagrangian approach) in two aspects:
the inclusion of processes with $K$ and $K^*$ in the initial and final states 
and the inclusion of processes with $Z_c(3900)$ and $Z_c(4025)$ in the intermediate states. 
We conclude that the interactions between $J/\psi$ and  all the 
considered  mesons reduce the original $J/\psi$ abundance ( 
determined at the end of the quark gluon plasma phase) by  20 \% and 
24 \% 
in RHIC and LHC  collisions respectively.  Consequently, any really   
significant change in the $J/\psi$  abundance comes from  dissociation and  
regeneration processes in the QGP phase. More details can be found in 
\cite{nos}.

\begin{acknowledgments}
This work was  partially financed by the Brazilian funding agencies 
CNPq, CAPES, FAPERGS and FAPESP. 
\end{acknowledgments}

\hspace{1.0cm}


\begin{thebibliography}{99}


\bibitem{qgp} J.~Schukraft, Nucl. Phys. A {\bf 967}, 1 (2017) and references 
              therein. 


\bibitem{review} M.~Nielsen and F.~S.~Navarra,  
                 Mod.\ Phys.\ Lett.\ A {\bf 29}, 1430005 (2014). 


\bibitem{brazzi} F.~Brazzi, B.~Grinstein, F.~Piccinini, A.~D.~Polosa and C.~Sabelli,
                 Phys.\ Rev.\ D {\bf 84}, 014003 (2011).



\bibitem{nos}    L.~M.~Abreu, K.~P.~Khemchandani, A.~Martinez Torres, 
                 F.~S.~Navarra and M.~Nielsen,
                 Phys.\ Rev.\ C {\bf 97}, 044902 (2018).


\bibitem{lee}    S. G. Matinyan and B. Mueller, Phys. Rev. C 58, 2994 (1998);  
                 Y. Oh, T. Song and S. H. Lee, Phys. Rev. C 63, 034901 (2001);  
                 K. L. Haglin and C. Gale, Phys. Rev. C 63, 065201 (2001); 
                 K. L. Haglin, Phys. Rev. C 61, 031902 (2000).



\bibitem{ff}  M.~E.~Bracco, M.~Chiapparini, F.~S.~Navarra and M.~Nielsen, 
              Prog.\ Part.\ Nucl.\ Phys.\  {\bf 67}, 1019 (2012);        
              F.~S.~Navarra, M.~Nielsen and M.~E.~Bracco,  
              Phys.\ Rev.\ D {\bf 65}, 037502 (2002). 

\bibitem{exhic} S. Cho et al. (ExHIC Collaboration), 
                Prog. Part. Nucl. Phys. {\bf 95}, 279 (2017).



\end{thebibliography}
\end{document}